# A Survey on Space-Time Turbo Codes

S.Nagarani, Research Scholar, Anna University, Coimbatore, Tamilnadu,India.
sureshnagarani@yahoo.co.in

Dr.C.V. Seshaiah, Prof and Head, Sri Ramakrishna Engg. College, Tamilnadu, India.
cvseshaiah@gmail.com

*Abstract*—As wireless communication systems look intently to compose the transition from voice communication to interactive Internet data, achieving higher bit rates becomes both increasingly desirable and challenging. Space-time coding (STC) is a communications technique for wireless systems that inhabit multiple transmit antennas and single or multiple receive antennas. Space-time codes make use of advantage of both the spatial diversity provided by multiple antennas and the temporal diversity available with time-varying fading. Space-time codes can be divided into block codes and trellis codes. Space–time trellis coding merges signal processing at the receiver with coding techniques appropriate to multiple transmit antennas. The advantages of space-time codes (STC) make it extremely remarkable for high-rate wireless applications. Initial STC research efforts focused on narrowband flat-fading channels. The decoding complexity of Space-time turbo codes STTC increases exponentially as a function of the diversity level and transmission rate. This proposed paper provides an over view on various techniques used for the design of space-time turbo codes. This paper also discusses the techniques handled by researchers to built encoder and decoder section for multiple transmits and receives antennas. In addition the future enhancement gives a general idea for improvement and development of various codes which will involve implementing viterbi decoder with soft decoding in a multi-antenna scenario. In addition the space-time code may be analyzed using some of the available metrics and finally to simulate it for different receive antenna configurations.

*Keywords*—Antenna, Bit Rate, Block Codes, Diversity, Space-Time Coding (STC), Transmitter, Trellis Codes, and Viterbi Decoder.

## I. INTRODUCTION

In general the Space-time codes utilize the advantages of both the spatial diversity provided by multiple antennas and the temporal diversity available with time-varying fading. Space-time coding (STC) is a communications technique for wireless systems that occupy multiple transmit antennas and single or multiple receive antennas. Therefore recently there has been a surge of interests in the design of the so called "space-time codes." Most of the proposed space-time codes were mainly based on trellis codes or orthogonal block codes. Those codes make use of a fixed number of transmitter antennas for each code construction. A new class of scalable space-time codes based on turbo codes or turbo trellis codes was proposed in [1]. They will be referred as space-time turbo codes (STT) in the consequence. The scalability means that the code rate and number of transmitter antennas can be effortlessly personalized to different design requirements without redesigning the main part of the code [26].

With high probability, this class of codes will take full advantage of both space and temporal diversity. The performance of the STT code is close or better than AT&T's 64 states code. Design criteria have been proposed for space-time codes in a assortment of environments [2], [3] and [4]. In Rayleigh fading, each feasible code word difference in a coded modulation produces a 'signal' matrix which is a function of both the code word difference and the channel correlation. The diversity provided by a code word difference is given by the rank of this 'signal' matrix and the effective product measure [4]. Therefore the STT model can utilize random symbol interleavers, such that the code word differences are stretched among all the transmitter antennas, which results in full rank and outsized effectual product measure 'signal' matrix with high probability.

The following advantages of space-time codes (STC) make it highly attractive for high-rate wireless applications. The foremost advantage is that it improves the downlink performance without the requirements for multiple receive antennas at the terminals. Second, it elegantly combines spatial transmit diversity with channel coding realizing a coding gains in addition to maximum diversity gain [3]. Moreover there is no necessity of channel state information (CSI) at the transmitter terminal. In addition the need for an expensive reverse link, which may be consistent in the case of rapid channel fading, can be eliminated by operating open loop. STC is robust against certain non-ideal operating conditions like antenna correlation, channel estimation errors and Doppler effects [5].

Multiple antennas posses the potential to drastically increase the achievable bit rates [6], thus converting wireless channels from narrow to wide data pipes. This can be well demonstrated using Information Theory. Rayleigh flat fading can be beneficial for a multiple-antenna communication link. More recently, space–time trellis coding has been proposed which combines signal processing at the receiver with coding techniques appropriate to multiple transmit antennas and provides significant gain over. This proposed paper provides an over view on various techniques used for the design of space-time turbo codes. In addition the future enhancement gives a general idea for improvement and development of various codes which will involve implementing viterbi decoder with soft decoding in a multi-antenna scenario. In addition the space-time code may be analyzed using some of the available metrics and finally to simulate it for different receive antenna configurations.

The remainder of this paper is structured as follows. Section 2 discusses the related work that was earlier proposed in literature for space-time turbo codes. Section 3 gives a general idea of further improvements of the earlier approaches in modeling the different sections of a space-time coding

## II. BACKGROUND STUDY

Space-Time Coding (STC) has been studied from early time and lots of advanced techniques have been proposed for implementing this communication technique for wireless systems. This section of the paper discusses the related work that was earlier proposed in literature for space-time turbo coding.

Alamouti in [7] proposed a simple two branch diversity scheme. The diversity created by the transmitter utilizes space diversity and either time or frequency diversity. Space diversity is affected by redundantly transmitting over a plurality of antennas, time diversity is affected by redundantly transmitting at different times, and frequency diversity is affected by redundantly transmitting at different frequencies. The scheme makes use of two transmitter antennas and one receiver antenna. Even then the proposed scheme provides the same diversity order as maximal-ratio receiver combining (MRRC) with one transmit antenna, and two receive antennas. The principles of this invention are applicable to arrangements with more than two antennas, (i.e. similarly it was proved) that the scheme can be generalized to two transmit antennas and M receive antennas, such that it may provide a diversity order of 2M. The most important advantage of the proposed scheme is that it does not require any bandwidth expansion or any feedback from the receiver to the transmitter. Additionally, the computational complexity of the proposed scheme is very much similar to MRRC.

Tarokh et al. in [8] described a space-time code that is applicable for high data rate wireless communications. Generally it is well known that Space-time coding is a bandwidth and power efficient method of communication over fading channels that realizes the remunerations of multiple transmit antennas. Precise codes have been constructed using design criteria consequent for quasi-static flat Rayleigh or Rician fading, where channel state information is accessible at the receiver. It is apparent that the reasonableness of space-time codes will be significantly improved if the derived design criteria continue to be applicable in the absence of perfect channel state information. It is even more enviable that the design criteria not be disproportionately sensitive to frequency selectivity and to the Doppler spread. They presented a theoretical study of these issues beginning with the effect of channel estimation error. They also assumed that the channel estimator extracts fade coefficients at the receiver and for constellations with constant energy, it is proved that in the absence of perfect channel state information the design criteria for space-time codes is still valid. They also derived the maximum-likelihood detection metric in the presence of channel estimation errors. They studied the effect of multiple paths on the performance of space-time codes for a slow changing Rayleigh channel. It is proved that the presence of multiple paths does not decrease the diversity order guaranteed by the design criteria used to construct the space-time codes.

A new paradigm for communication was introduced by Tarokh et al. in [9]. They introduced a space-time block coding, a new paradigm for communication over Rayleigh fading channels using multiple transmit antennas. Data is encoded with the aid of a space-time block code and the encoded data is divide into 'n' streams which are concurrently transmitted using 'n' transmit antennas. The received signal at each receive antenna is a linear superposition of the 'n' transmitted signals disconcerted by noise. Maximum-likelihood decoding was accomplished in an uncomplicated means through decoupling of the signals transmitted from different antennas rather than joint detection. This uses the orthogonal structure of the space-time block code and gives a maximum-likelihood decoding algorithm which is based only on linear processing at the receiver. The approach focuses on achieving the maximum diversity order for a provided number of transmit and receive antennas subject to the limitation of having a simple decoding algorithm. The space-time block code is constructed using the classical mathematical framework of orthogonal designs. Consequently, a generalization of orthogonal designs is shown to afford space-time block codes for both real and complex constellations for any number of transmit antennas as the code constructed in the above way exist only for a few sporadic values of 'n'. These codes realize the greatest possible transmission rate for any number of transmit antennas using any uninformed real constellation such as Pulse Amplitude Modulation (PAM). The best tradeoff between the decoding delay and the number of transmit antennas was also computed and they showed that many of the codes presented here are optimal in this sense as well.

Tarokh et al. in [10] put forth a practical way to illustrate that the information capacity of wireless communication systems can be increased dramatically by employing multiple transmit and receive antennas. An effective approach to increasing data rate over wireless channels is to employ space-time coding techniques suitable to multiple transmit antennas. These space-time codes initiate sequential and spatial correlation into signals transmitted from different antennas, so as to provide diversity at the receiver, and coding gain over an un-coded system. Their proposed approach noticeably reduced encoding and decoding complexity. This was achieved by partitioning antennas at the transmitter into small groups, and using individual space-time codes, called the component codes, to transmit information from each group of antennas. A novel linear processing which is capable of suppressing the signals transmitted by other group of antennas by treating them as interference was employed at the receive antenna to decode an individual space-time code. The simple receiver structure provides the diversity and the coding gain over un-coded system. This combination of array processing at the receiver and coding techniques for multiple transmit antennas can offer steadfast and very high data rate communication over narrowband wireless channels. A modification of this fundamental configuration gives rise to a multilayered space-time structural design that both generalizes and improves upon the layered space-time architecture.

The performance of space–time block codes which provided a new standard for transmission over Rayleigh fading channels using multiple transmit antennas was documented by Tarokh et al. in [11]. They considered a wireless communication system with 'n' antennas at the base station and 'm' antennas at the remote. The main purpose of their paper is to estimate the performance of the space–time block codes constructed them in their earlier work and to provide the details of the encoding and decoding procedures. They assumed that transmission at the base-band employs a signal constellation. Maximum likelihood decoding of any space–time block code can be achieved using only linear processing at the receiver. Figure 1 shows the system block diagram of space-time coding.

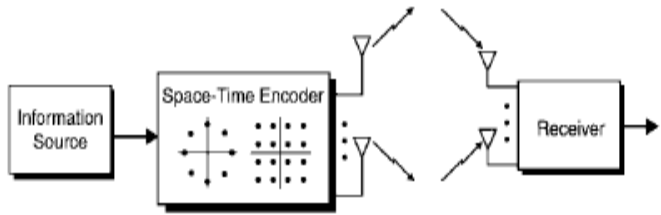

Figure.1. System block diagram of STC Scheme

The information source is encoded using a space–time block code, and the constellation symbols are transmitted from different antennas. The receiver estimates the transmitted bits by using the signals of the received antennas. Their experimental results revealed the fact that considerable gains can be achieved by increasing the number of transmit antennas with very little decoding complexity. In addition there exists a possibility to concatenate an outer trellis code [12] with the above space–time block coding to achieve even better performance. The added coding gain provided by the outer code is the same as the gain provided by that code on a Gaussian channel. The decision metrics given in their paper for the inner space–time block code then can be used as the branch metrics for the outer trellis code [12]. This improves the performance at the disbursement of a higher complexity.

Sandhu and paulraj in [13] presented the Space-time block codes as a remarkable modulation scheme discovered recently for the multiple antenna wireless channels. They have a well-designed mathematical solution for providing full diversity over the coherent, flat-fading channel. In addition, they necessitate extremely simple encoding and decoding at the transmit antenna and the receive antenna respectively. They showed that even though the scheme provided full diversity at low computational costs the scheme incur a loss in its capacity. This is because they convert the matrix channel into a scalar AWGN channel whose capacity is smaller than the true channel capacity. In their view the loss in capacity is quantified as a function of channel rank, code rate, and number of receive antennas.

Telatar in [14] investigated the use of multiple transmitting and/or receiving antennas for single user communications over the additive Gaussian channel with and without fading. Formulas for the capacities and error exponents of such channels were also derived and evaluated the computation procedures for those formulas in [14]. Additionally, the paper also revealed the potential gains of such multi-antenna systems over single-antenna systems is rather large under independence assumptions for the fades and noises at different receiving antennas. A single user Gaussian channel with multiple transmitting and/or receiving antennas has been taken into account for deriving the formulas and the computation procedures. This choice models a Rayleigh fading environment with enough separation within the receiving antennas and the transmitting antennas such that the fades for each transmitting-receiving antenna pair are independent. Error components provide an answer for the hardness to get closer to the capacity of the system by giving an upper bound to the probability of error achievable by block codes of a given length 'n' and rate 'R'. The upper bound is known as random coding bound. The multi-user capacity for this communication scenario can be evaluated easily by exploiting the nature of the solution to the single user scenario. The use of multiple antennas will significantly augment the attainable rates on fading channels if the channel parameters can be estimated at the receiver and if the path gains between different antenna pairs behave independently.

Foschini et al. in [15] presented a paper that was greatly motivated by the need for fundamental understanding of ultimate limits of bandwidth efficient delivery of higher bit-rates in digital wireless communications and to also begin to look into how these limits might be approached. They examined the development of multi-element array (MEA) technology, which is processing the spatial dimension to improve wireless capacities in certain applications. Exclusively, they presented a quantity of essential information theory results that guarantee great advantages of using MEAs in wireless LANs and building to building wireless communication links. The case where the channel characteristics is not available at the transmitter but the receiver tracks the characteristic which is subject to Rayleigh fading has been explored in their presented paper. The capacity offered by MEA technology was revealed by fixing the over all transmitted power. They investigated the case of independent Rayleigh faded paths between antenna elements and find that with high probability extraordinary capacity is available. Standard approaches such as selection and optimum combining are seen to be incomplete when compared to what will eventually be possible. New codecs need to be invented to comprehend a robust portion of the great capacity promised.

A new point-to-point communication that employs an equal number of array elements at both the transmitter and the receiver was described by Foschini in [16]. The architecture is specially designed for a Rayleigh fading environment in situations in which the transmitter does not have the knowledge of the channel characteristic. The architecture is a method of presenting and processing higher dimensional signals with the focus of leveraging the highly developed one-dimensional codec technology. The capacity is achieved in terms of n equal lower component capacities one for each antenna at the receiver or at the transmitter. The following notations and fundamental assumptions should be reviewed Number of Antennas, transmitted signal, Noise at the Receiver, Received signal, Average Signal-to-noise ratio (SNR) at each receive antenna, and matrix channel impulse response. The random channel model used in their approach is Rayleigh channel model. It is assumed that the MEA elements at each end of the communication link are isolated by about half a wavelength. The complementary capacity distributions discussed by them mainly concentrates on high-probability tail. The distribution of capacity was derived from an ensemble of statistically independent Gaussian matrices. The central theme of their architecture is interference avoidance and it assumes that the interfering signals will be nulled out.

Wolniansky et al. in [17] proposed architecture for realizing very high data rates over the rich scattering wireless channels. They described a wireless communication architecture known as vertical BLAST (Bell Laboratories Layered Space-Time) or V-BLAST. The essential difference between D-BLAST and V-BLAST lies in the vector encoding process. In V-BLAST, however, the vector encoding progression is basically a de-multiplex operation followed by independent bit-to-symbol mapping of each substream. No inter-substream coding, or coding of any kind, is required, nevertheless conventional coding of the individual substreams may undoubtedly be applied. V-BLAST is essentially a single-user system which uses multiple transmitters. The following advantages of the BLAST system differs it from other conventional multiple access techniques in a single user fashion. The total channel bandwidth utilized in the BLAST system is only a small fraction in excess of the symbol rate, each transmitted signal occupies the entire system bandwidth, and the entire system bandwidth is used concurrently by all of the transmitters all of the time. V-BLAST utilizes a combination of old and new detection techniques to separate the signals in an efficient manner, permitting detection at significant fractions of the Shannon capacity and achieving large spectral efficiencies in the process. One way to execute detection for this system is by using conventional adaptive antenna array (AAA) techniques, i.e. linear combinatorial nulling. Superior performance can be achieved by using non-linear technique rather than a linear technique for detection of this system.

Golden et al. in [18] formulated a detection algorithm using V-BLAST space-time communication architecture. They described a simplified version of the BLAST detection algorithm, known as vertical BLAST, or V-BLAST, which has been implemented in real time in the laboratory. The detection process uses linear combinatorial nulling and symbol cancellation for computation. The V-BLAST system diagram is shown in the Figure 2.

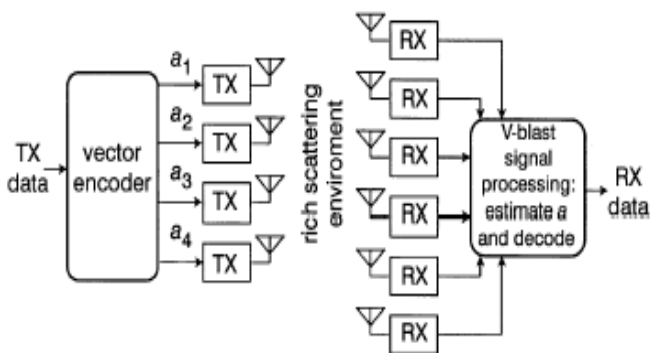

Figure.2 System Block diagram of V-BLAST Scheme

For simplicity in the sequel, they assumed that the same constellation is used for each component, and that transmissions are organized into bursts of L symbols. The power launched by each transmitter is proportional to 1/M so that the total radiated power is constant and independent of M, where M indicates the total number of QAM transmitters that operate co-channel at symbol rate 1/T symbols, with synchronized symbol timing. The receivers also operate co-channel, each receiving the signals radiated from all M transmit antennas. All their results were obtained in a short-range indoor environment with negligible delay spread.

A low density parity check codes for space application was designed and standardized by Andrews et al. in [19]. Their paper described recent activities in the next generation of FEC code development and standardization, which has concentrated on low-density parity-check (LDPC) codes. These codes also achieve excellent performance, within about 1 dB of the Shannon limit. Bandwidth efficiency is largely achieved through modulation design, but it affects code selection through the code rate. Decoders for LDPC codes use a message-passing algorithm based on the Tanner graph. A simple computation is performed at each variable node to compute soft code symbol estimates, these estimates are passed as messages to the check nodes where more simple computations are performed to compute correction terms, and these correction terms are passed back to the variable nodes. Modern turbo and LDPC codes considerably outperform traditional Reed-Solomon, convolutional, and other codes, and specific codes have been optimized for the constraints of spacecraft use.

Burr in [20] considered the application of space-time processing techniques using multiple antennas to third generation mobile communication systems, which have been shown to enable substantial capacity improvements. A technique was described for the evaluation of such schemes and apply it to some potential applications of the technique within the current third generation standards. The application also included a simple technique which can provide a 16th order diversity improvement, resulting in a 10 dB performance improvement. Both the dependence of the capacity on the radio environment, and the requirements of signal design for these channels were considered. These applications can be implemented to third generation mobile communication systems to enable substantial capacity improvements. Multi-user detection is also important to provide a good signal to noise ratio.

Silva et al. in [21] put forth a frequency domain receiver for rate 1 space-time block codes. Their paper considers iterative frequency-domain receivers for block transmission techniques with rate-1 Space Time Block Coding (STBC) for two and four transmit antennas using both Orthogonal Frequency Division Multiplexing (OFDM) and Single-Carrier (SC) schemes. Their proposed receiver includes an interference canceller which enhances the performance of the non-orthogonal STBC scheme with 4 transmit antennas, allowing performances close to those of orthogonal codes. Additionally their performance results show that combining STBC with block transmission techniques allows excellent performances.

Unitary space-time modulation scheme was proposed by Hochwald et al. in [22]. They proposed a signaling scheme, unitary space-time modulation, for multiple-antenna communication links. This modulation is preferably appropriate for Rayleigh fast-fading environments, since it does not necessitate the receiver to know or learn the propagation coefficients. Unitary space-time modulation uses constellations of T×M space-time signals where T represents the coherence interval during which the fading is approximately constant and M<T is the number of transmitter antennas. When the receiver does not know the propagation coefficients, which between pairs of transmitter and receiver antennas are modeled as statistically independent, this modulation performs very well when the signal-to-noise ratio (SNR) is high. They designed some multiple-antenna signal constellations and simulated their effectiveness as measured by bit-error probability with maximum-likelihood decoding.

A quasi random approach for to space-time codes was described by Wu et al. in [23]. The fundamental principle is to transmit randomly interleaved versions of forward error correction (FEC)-coded sequences concurrently from all antennas in a multilayer structure. In their proposed paper they provided a comprehensive study for the interleave-division-multiplexing space–time (IDM-ST) scheme. Additionally they developed a technique evaluating the signal-to-noise ratio (SNR) that is closely related to the extrinsic information transfer (EXIT) chart method. They examined quite a lot of power allocation strategies to maximize power efficiency as well as spectral efficiency using the fast performance assessment technique. They developed an equalization technique for IDM-ST codes in ISI channels following the linear minimum mean square error (LMMSE) and probabilistic data association (PDA) principles. This technique provides a simple, fast, and relatively accurate technique for performance evaluation.

Baro et al. in [24] put forth an approach for improving the performance of BLAST. The error propagation limits the practical implementation of BLAST architecture. The theoretically possible increase in diversity level during successive detection steps cannot be achieved. They evaluated the possibilities to improve the performance of the earlier proposed BLAST system. They initially analyzed an MMSE solution compared to zero forcing. Besides, they had showed the benefits of soft over hard interference cancellation and proposed an iterative turbo detection algorithm. As a final point, they employed transmit diversity with space-time block codes to BLAST. This improves the overall diversity level and reduces the number of required receive antennas.

Al-Dhahir et al. in [25] presented an overview of research activities on space-time coding for broadband wireless transmission performed at AT&T Shannon Laboratory over the past two years. Their primary emphasis is on physical layer modem algorithms such as channel estimation, equalization, and interference cancellation. On the other hand, they also discussed the collision of space-time coding gains at the physical layer on throughput at or above the networking layer. In addition, they also described a flexible graphical user interface attached to our physical layer simulation engine in order to investigate the performance of space-time codes under an assortment of practical transmission scenarios. Their work also presented the simulation results for the EDGE cellular system and the 802.11 wireless Local Area Network (LAN) environments.

A recursive space-time trellis codes was projected by Fu et al. in [27] using differential encoding. In their paper, they also revealed their interest in pursuing large diversity product space–time codes. The major difference between the above systematic scheme and the traditional scheme is that the states in their scheme are generated by a non-group unitary space–time code (USTC) while the states in the latter are from a group USTC itself and a group USTC limits the number of states and its diversity product that may affect the inner code performance. They proposed a new USTC design criterion to ensure that the trellis structure improves the diversity product over the USTC as a block code. They put forth a new class of USTC design for an arbitrary number of transmit antennas that has an analytical diversity product formula for two transmit antennas.

## III. Future Enhancement

This section of the paper explains further improvements of the earlier approaches in modeling the different sections of a space-time coding scheme. The future works rely on studying the code and analyze it using the metrics developed in [8]. The experiments may be carried out on different receive antenna configuration. The space-time code may be analyzed and simulated by implementing viterbi decoder with soft decoding in a multi-antenna scenario. The performance of the linear receivers like Zero Forcing (ZF) receiver, Minimum Mean Squared Error (MMSE) receiver, will analyzed for space-time coding. Finally, the future work may focus on successive interference cancellation receiver known as Vertical Bell Laboratories Layered Space-Time Architecture (V-BLAST). The performance of V-BLAST receiver may be analyzed by comparing it with other linear receivers. Spatial multiplexing techniques send independent data streams on different transmit antennas to maximally exploit the capacity of multiple-input multiple-output (MIMO) fading channels. Most existing multiplexing techniques are based on an idealized MIMO channel model representing a rich scattering environment. Simulations for both space-time coding and spatial multiplexing [28] will be done using Monte Carlo method via Lab View. The channel model will be assumed to be a Rayleigh fading model and will be constant over a block of symbols. The performance of the space-time coding can be evaluated under different antenna configurations.

## IV. Conclusion

Space-time coding (STC) is a technique for wireless systems that reside in multiple transmit antennas and single or multiple receive antennas. Most of the proposed space-time codes were mainly based on trellis codes or orthogonal block codes. This proposed paper provides an over view on a variety of techniques used for the design of space-time turbo codes. In addition the future enhancement gives a wide-ranging suggestion for improvement and development of a range of codes which will involve implementing viterbi decoder with soft decoding in a multi-antenna scenario. In addition the space-time code may be analyzed using some of the available metrics and finally to simulate it for different receive antenna configurations. In addition the space-time code may be analyzed using some of the available metrics and finally to simulate it for different receive antenna configurations.

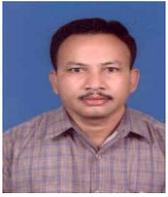

Dr. C. V. Seshaiah received his M.Sc., in Mathematics from S. V. University, Tirupati, Andhra Pradesh, in the year 1982. He obtained his Ph.D., in Mathematics from Bhavnagar University, Gujarat, in the year 1990. Currently, he is working as a professor and the Head of the Department (Mathematics) in Sri Ramakrishna Engineering College, Coimbatore, and Tamilnadu. He has a total of 24 years work experience. His areas of interest include general theory of relativity, mathematical statistics, and commutative algebra. He has published eight research papers in national and international journals. He had published two books on engineering mathematics. He has presented five papers in national and international seminars held at various institutions.

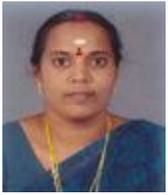

S. Nagarani received her B.Sc., and M.Sc., in Mathematics from Avinashilingam Deemed University, Coimbatore, and Government Arts College affiliated to Bharathiar University, Coimbatore, in the year 1994 and 1996 respectively. She received her M.Phil., in Fluid Dynamics from Bharathiar University, Coimbatore in 1998. Currently, she is pursuing her Ph.D., in Anna University, Coimbatore, and Tamilnadu. She has a work experience of 12 years. At present she is working as a senior lecturer in the department of Mathematics. She has published papers in national and international journals. She has participated and presented research papers in various national and international seminars and conferences. She has been one of the reviewers of Engineering Mathematics book for Pearson's Publications.